\documentclass{aa}  
\usepackage{graphicx}
\usepackage{txfonts}
\usepackage[]{hyperref}
\usepackage{xcolor}
\usepackage[normalem]{ulem}

\urldef{\sunkitspexbasic}\url{https://github.com/sunpy/sunkit-spex}
\urldef{\sunkitspexstix}\url{https://github.com/natsat0919/sunxspex/tree/albedo-correction}
\urldef{\stixdatacenter}\url{https://datacenter.stix.i4ds.net/}
\urldef{\stixdemo}\url{https://sunkit-spex.readthedocs.io/en/latest/generated/gallery/fitting_STIX_spectra.html}

\begin{document} 

   \title{Superhot (> 30 MK) flare observations with STIX: Joint spectral fitting}

   \author{Muriel Zo\"{e} Stiefel
          \inst{1}\fnmsep\inst{2}
          \and
          Nat{\'a}lia Bajnokov{\'a} \inst{3}
          \and
          S\"{a}m Krucker\inst{1}\fnmsep\inst{4}
          }

   \institute{University of Applied Sciences and Arts Northwestern Switzerland, Bahnhofstrasse 6, 5210 Windisch, Switzerland\\
              \email{muriel.stiefel@fhnw.ch} 
        \and
            Institute for Particle Physics and Astrophysics, ETH Z\"{u}rich, 
            R\"{a}mistrasse 101, 8092 Z\"{u}rich Switzerland
        \and
            School of Physics \& Astronomy, University of Glasgow, University Avenue, Glasgow G12 8QQ, UK
        \and
            Space Sciences Laboratory, University of California, 7 Gauss Way, 94720 Berkeley, USA
            }

   \date{Received June 12, 2025; accepted November xx, 2025}
 
  \abstract
   {Spectroscopic analysis of large flares (>X1) in the hard X-ray (HXR) range offers unique insights into the hottest (> 30 MK) flare plasma, the so-called superhot thermal component. To manage the high count rates in large flares, an attenuator is typically placed in front of the HXR detectors. However, this significantly limits the spectral diagnostic capabilities at lower energies, and consequently, it restricts the analysis of the lower temperatures in flares.}
   {The Spectrometer/Telescope for Imaging X-rays (STIX) on board the Solar Orbiter mission was designed to observe solar flares in hard X-rays. The imaging detectors use an attenuator during periods of high flux level. In contrast, the background (BKG) detector of STIX is never covered by the attenuator and is therefore dedicated to measure the unattenuated flux using differently sized apertures placed in front of the detector. We aim to demonstrate that joint spectral fitting using different detector configurations of STIX allows us to reliably diagnose both the hot and the superhot components in large flares.}
   {We jointly fit the HXR spectra of the STIX BKG detector and the STIX imaging detectors using SUNKIT-SPEX software package to determine the spectral parameters of both the hot and superhot thermal components in solar flares.}
   {Using joint fitting on 32 STIX flares, we corroborate that for GOES X-class flares, the HXR spectrum is better represented by two thermal components instead of an isothermal component. At the temperature peak time, the superhot HXR flux above $\sim$15 keV is typically stronger than the hot HXR flux. The GOES long-wavelength channel is dominated by the hot component with a superhot contribution up to 10\%.}
   {This paper demonstrates that joint spectral fitting of the same detector type with different attenuation schemes is a simple and powerful method to monitor multithermal flare plasma.}


   \maketitle

\section{Introduction}\label{Intro}

    Solar flares are the most energetic events in our solar system. They are detectable across the entire electromagnetic spectrum, even producing high-energy emissions in the hard X-ray (HXR) and gamma-ray range \citep[for a review e.g., ][]{Benz_2017}. The first HXR spectrum linked to a solar flare was reported by \citet{Peterson_1958} during a cosmic-ray balloon flight. The emission was interpreted as bremsstrahlung produced by flare-accelerated nonthermal electrons hitting the solar chromosphere. Subsequently, several balloon and rocket missions reported soft X-ray (SXR) and HXR emissions from the Sun associated with solar flares. Due to the coarse energy resolution of the measurements, a debate arose whether the observed spectra could be explained by nonthermal emission \citep[e.g., ][]{Peterson_1959, Anderson_1962}, thermal emission \citep[e.g., ][]{Chubb_1966, Snijders_1969}, or a combination of both \citep[e.g., ][]{Acton_1968, Takakura_1969}. The first high-resolution spectral measurement of a HXR burst was reported by \citet{Lin_1981} during a balloon mission. This measurement provided an answer on which model to use for spectral analysis. At high energies (above $\sim$ 30 keV), the spectrum was best fitted by a power-law fit \citep[e.g., thick-target model ][]{Brown_1971}. At lower energies, the spectrum exhibited a much steeper power-law ($\gamma \gtrapprox 11$), which was better represented by a single thermal Maxwellian electron distribution (isothermal model). Today, the combination of both a nonthermal and an isothermal model is still commonly used in spectral analysis within high-energy solar physics. In terms of flare models, the nonthermal model reflects the emission of nonthermal electrons that were accelerated in the corona due to magnetic reconnection and hit the denser chromosphere, where they produce bremsstrahlung. The interaction of nonthermal electrons with the chromosphere heats the plasma, which leads to chromospheric evaporation \citep[e.g., ][]{Antonucci_1982}, building the hot flare loops. The hot flare loops produce bremsstrahlung that is detected by HXR instruments as thermal emission in the lower energy range.

    Further, in the study by \citet{Lin_1981}, they reported a new spectral component in HXR. A much hotter thermal component ($\sim$ 34 MK) than previously reported temperatures from solar flares and hotter than the GOES SXR temperature of the flare studied. This new component was subsequently confirmed to be of thermal origin \citep{Emslie_1989} and was named the superhot component \citep[e.g., ][]{Hudson_1985}.

    In 2002, the Reuven Ramaty High Energy Solar Spectroscopic Imager \citep[RHESSI;][]{Lin_2002} was launched. This spacecraft was designed to perform high spectral and spatial resolution observations of HXR down to $\sim$ 3 keV, making it sensitive to plasma temperatures $\gtrsim$ 10 MK. In a detailed analysis of an X4.8 GOES class flare, \citet{Caspi_2010} demonstrated that the spectral fitting could be improved for a flare of this magnitude by using a nonthermal model and two thermal models instead of just one. These two thermal components represent plasma populations in different temperature regimes. The first, hot component corresponds to lower temperatures, typically ranging from $\sim$15-30 MK and aligning with the derived GOES temperatures. The second, superhot component represents temperatures exceeding 30 MK, comparable to the superhot temperatures reported by \citet{Lin_1981}. \citet{Caspi_2014} found that most large flares ($\gtrsim$ X1) have a superhot thermal component. On the other hand, there are instances of smaller X-class flares with no distinct superhot thermal component and overall cooler temperatures \citep[e.g., ][]{Caspi_2014b}.
    
    Despite these results, fitting the thermal component using a HXR instrument has limitations due to the poor sensitivity to cooler plasma (i.e., low energy emission). The uncertainty is even more enhanced for large flares as HXR instruments typically utilize attenuators to mitigate pileup (incoming photons during the readout time distort the recorded photon energy) and live-time concerns (decrease of the recording time for the incoming flux due to the read-out time of the detectors for an event), and to maintain sensitivity to the nonthermal emission. In the case of RHESSI, the attenuator was not uniformly thick; instead, a region in the middle of the attenuator was thinner \citep{Smith_2002}. This design allowed for some low-energy flux to be detectable, although the calibration is more complicated. \citet{Caspi_2010} showed that the method is effective for attenuated flares; nevertheless, the fitting of the thermal component does not always give unique solutions. Another way is to combine a HXR instrument with a SXR instrument. This was done by \citet{Nagasawa_2022}, where they used the Miniature X-ray Solar Spectrometer \citep[MinXSS;][]{Mason_2016} alongside RHESSI and jointly fitted the two spectra. MinXSS is sensitive to lower energies, which helps in the thermal fitting of the hot component. In contrast, RHESSI observes higher energies, including the superhot thermal component and the nonthermal component. 

    In 2020, Solar Orbiter was launched \citep{Muller_2020}. Onboard is the Spectrometer/Telescope for Imaging X-rays \citep[STIX;][]{Krucker_2020}, which is designed to observe HXR in the energy range of 4-150 keV. This makes STIX sensitive to plasma temperatures $\gtrsim$ 10 MK. Similar to RHESSI, STIX utilizes an attenuator that is automatically inserted when a certain trigger threshold is reached. The attenuator blocks the lower energy flux during significant flare events. STIX also includes a detector known as the background (BKG) detector, which is never covered by the attenuator \citep{Stiefel_2025}. The BKG detector provides unattenuated spectral observations in the lower energy channels of STIX using three different aperture sizes placed in front of the detector. Consequently, the BKG detector can take over a similar role during attenuated flares of STIX as MinXSS did in \citet{Nagasawa_2022}, with the advantage that the BKG detector uses the same detector type as the other detectors of STIX. The spectrum of the imaging detectors of STIX is primarily dominated by the superhot and nonthermal components, whereas the BKG detector spectrum is mainly dominated by the hot thermal component.
    
    In \citet{Stiefel_2025}, an iterative method was presented for fitting the spectrum of the BKG detector together with the spectra obtained by the imaging detectors of STIX using OSPEX, the standard tool to date used for spectral fitting with STIX data. In this paper, we introduce an improved method compared to \citet{Stiefel_2025} by jointly fitting the two spectra simultaneously using SUNKIT-SPEX. This approach is using a similar method as discussed in \citet{Bajnokov_2024}, where STIX spectra were jointly fitted with Nuclear Spectroscopic Telescope ARray \citep[NuSTAR; ][]{Harrison_2013} spectra. 

    The structure of this paper is as follows: We begin with a more detailed description of STIX in Section \ref{Chapter: Data}. In Section \ref{Chapter: Method}, we outline the method of joint fitting between the two STIX spectra. In Section \ref{Chapter: Application}, we apply the new method to 32 large STIX flares, comparing our results with the existing literature. Finally, we conclude and discuss the optimal use of attenuators for HXR instruments in future space missions.

\section{STIX HXR observations} \label{Chapter: Data} 

    STIX is a spectrometer/telescope measuring solar flares in the energy range of 4-150 keV \citep{Krucker_2020}. For imaging, STIX employs an indirect imaging system by measuring 30 Fourier components, called 'visibilities', with 30 detectors. The visibilities are then used to reconstruct images \citep[for more details, see ][]{Massa_2023}. When mentioning the imaging detectors in this paper, we are always referring to the 24 detectors with the coarsest grids, as they are less sensitive to variations in the grid calibration. STIX records spectral information by measuring the energy of the incoming photon and allocating it to an energy bin, with the finest energy resolution of 1 keV bins in the lower energy range. To minimize pileup or live-time issues with the detectors and to ensure the measurement of the nonthermal emission, an attenuator moves in when reaching a certain trigger threshold. When the attenuator is inserted, the energy ranges of 4-12 keV from STIX should not be used for imaging and spectral analysis due to an increased contribution of scattered and secondary photons. Therefore, part of the information on the thermal emission is missing for attenuated flares. 
    
    The BKG detector is the only detector of STIX that is not covered by the attenuator. As a result, it measures unattenuated low-energy flux during the largest solar flares. The BKG detector is mostly shielded by a tungsten plate, allowing solar flux to reach the detector only through six different-sized openings (two openings have the same size for redundancy), which are each placed directly in front of one pixel. Since the BKG detector lacks an attenuator, individual pixels are turned off during periods of high flux to prevent issues related to pileup and detector live-time \citep[for more details on the design, refer to ][]{Stiefel_2025}. 

\section{Methodology} \label{Chapter: Method}

    The spectral fitting in this study is done using SUNKIT-SPEX\footnote{\sunkitspexbasic}, the SunPy solar X-ray fitting package \citep{sunpy_2022}. SUNKIT-SPEX uses the following approach for fitting: A parametric photon model is assumed and convolved with the spectral response matrix (SRM). This convolved model is then compared to the observed data. Through forward fitting, the model that best fits the data is identified by maximizing the negative log-likelihood using a Bayesian approach. Traditionally, OSPEX (Object Spectral Executive) is used for spectral fitting with STIX data. However, one advantage of SUNKIT-SPEX over OSPEX is its ability to fit multiple spectra simultaneously \citep[e.g., ][]{Cooper_2024, Bajnokov_2024}. Additionally, SUNKIT-SPEX employs a Bayesian approach for fitting, enabling Markov chain Monte Carlo (MCMC) analysis on the fit to explore the model parameter space.  

    \citet{Bajnokov_2024} used the SUNKIT-SPEX package for the first time with STIX data by jointly fitting STIX and NuSTAR spectra. In this paper, we used an updated version of the same package. In the following, we will outline the basic procedure for a joint fit using two STIX spectra. We will then discuss the advantages of using joint fitting for attenuated STIX flares. Finally, we will highlight a few advantages of this new method compared to the iterative approach presented in \citet{Stiefel_2025}.

\subsection{Joint fitting with STIX observations}

    For spectral analysis, the SUNKIT-SPEX package takes two input files; the spectrum fits file and the associated SRM fits file. To date, both files still need to be created in IDL using the Ground Software (GSW) of STIX. This paper uses the STIX GSW version v0.5.2. The spectrum and SRM files are created using the routine stx\_convert\_pixel\_data.pro in the GSW, along with the level 1 fits files downloaded from the STIX data center\footnote{\stixdatacenter \label{footnote: Data center}} \citep{Xiao_2023}. The background is subtracted from the spectrum using a STIX background file. The background files consist of measurements from the STIX detectors taken during non-flaring periods. For each flare, the background file that is closest in time is used in the analysis. For joint fitting, it is necessary to create the spectrum and the SRM files for both detector configurations. For the BKG detector, we only use the counts recorded by the pixels behind the middle sized openings, pixels 2 and 5. These pixels are selected for their optimal statistics on flare counts when the attenuator is inserted \citep[for more details, see ][]{Stiefel_2025}. For the imaging detectors, we use all large pixels. 

    We maximize the negative $\chi^2$ log-likelihood when fitting STIX data in SUNKIT-SPEX. In joint fitting, the photon model is tied to each spectrum simultaneously. Each spectrum is therefore contributing to the total likelihood individually. The 1$\sigma$ on the output parameters of the fitting model are taken from the MCMC analysis.

    \begin{figure}
    \centering
    \includegraphics[width=0.44\textwidth]{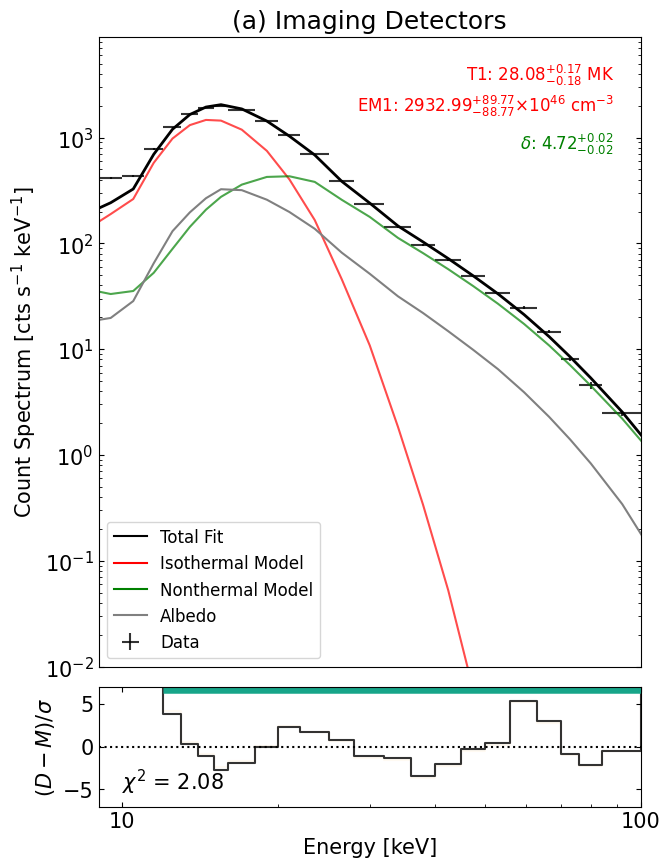}
    \includegraphics[width=0.44\textwidth]{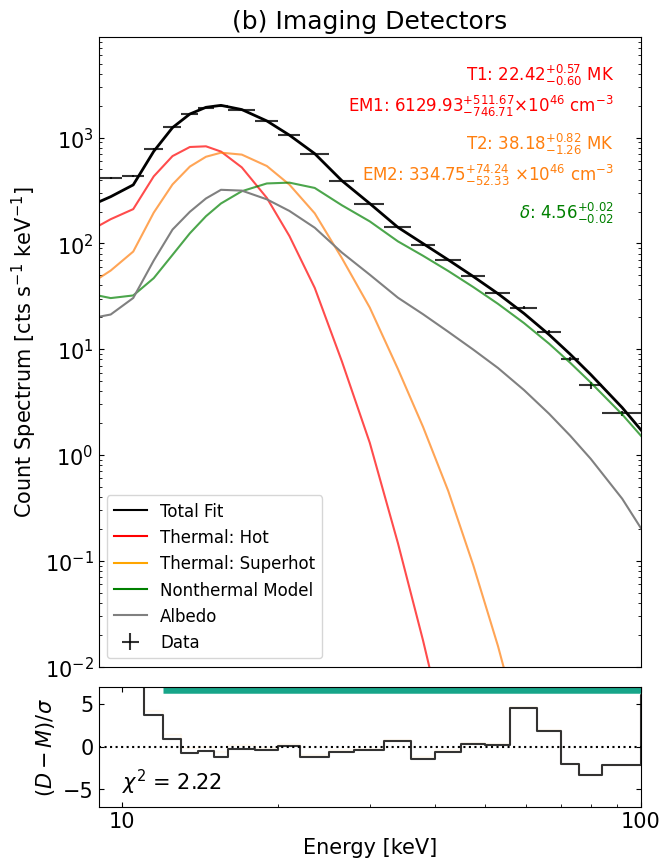}
    \caption{Spectrum of the estimate X5-class flare SOL20230103 recorded by the imaging detectors of STIX. The spectrum is integrated over a time range of 10 s between 06:26:50-06:27:00 UTC (Earth time). The measured spectrum is the same in both panels, only the fitting model is different: For the top panel, a photon model including an isothermal model (red), a nonthermal model (green), and an albedo component (grey) is fitted. The total fit is given in black. In the bottom panel, we used the same components for the photon model together with a second thermal component (orange). Below each spectrum, the residuals (data minus model divided by the uncertainty on the data points) and the $\chi^2$ are given. The green bar indicates the energy range used for fitting.}
    \label{Fig: Single Fit Imaging}
    \end{figure}

    \begin{figure*}
    \centering
    \includegraphics[width=0.80\textwidth]{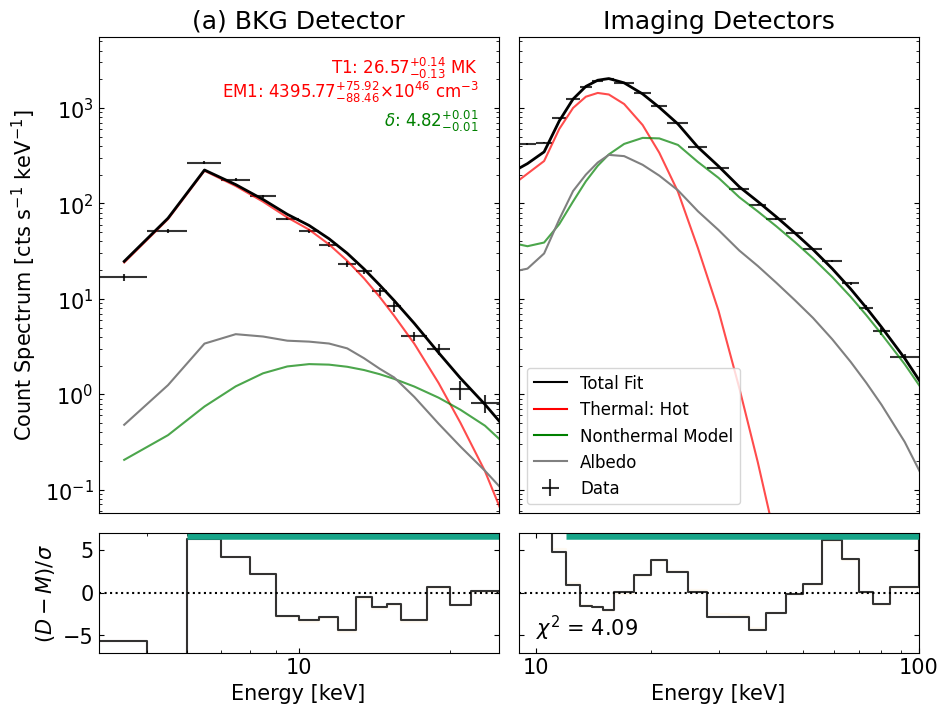}
    \includegraphics[width=0.80\textwidth]{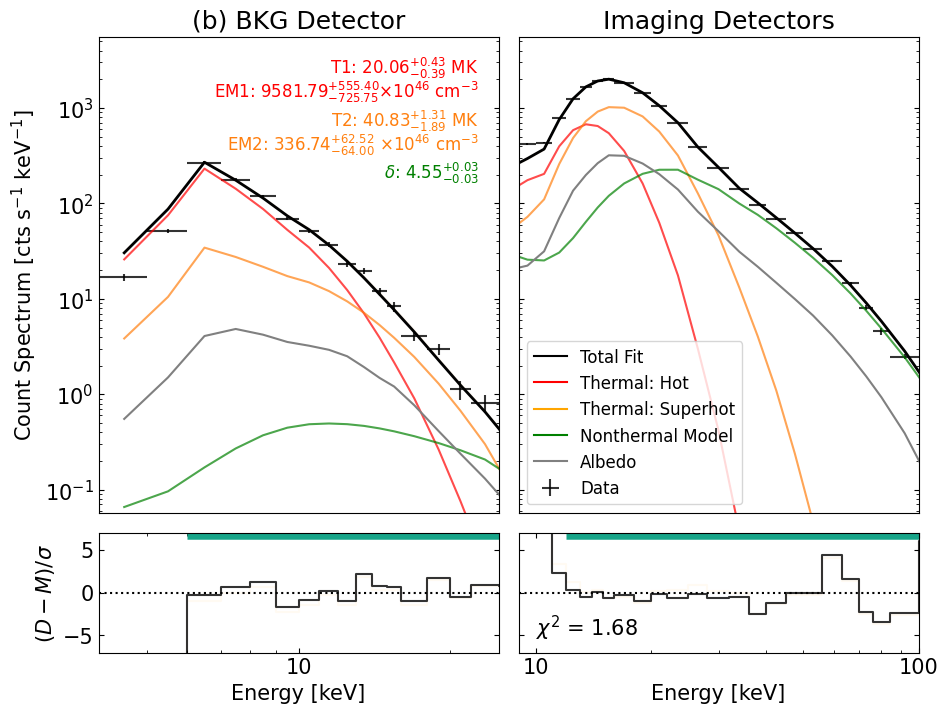}
    \caption{Spectrum of the estimate X5-class flare SOL20230103 measured by the BKG detector (left plots) and the imaging detectors (right plots, same as Fig. \ref{Fig: Single Fit Imaging}) of STIX. The spectrum is integrated over a time range of 10 s between 06:26:50-06:27:00 UTC (Earth time). The measured spectrum is the same in the top and bottom panels, only the fitting model is different: For the top panel, a photon model including an isothermal model (red), a nonthermal model (green), and an albedo component (grey) is jointly fitted to the two spectra. The total fit is given in black. In the bottom panel, we used the same components for the photon model together with a second thermal component (orange). Below all spectra, the residuals (data minus model divided by the uncertainty on the data points) and the $\chi^2$ are given. The green bar indicates the energy ranges used for fitting.}
    \label{Fig: Joint Fit}
    \end{figure*}
    
    For the photon model, we use combinations of three different models: The isothermal model (labeled with f\_vth in SUNKIT-SPEX), the nonthermal model (labeled with thick\_fn) based on the thick-target model \citep{Brown_1971}, and the albedo emission component (can be activated with "albedo\_corr = True"). The albedo component originates from photons that travel downwards into the solar atmosphere and are Compton back-scattered towards the observer \citep{Kontar_2006}. The strength of the albedo component depends on the angle between the normal vector on the solar surface at the flare location and the observer's position. To determine this angle, we use the STIX GSW routines to estimate the flare location, along with tools available on the STIX data center\footnotemark[2] to calculate the angle based on that location.  

    To take into account possible systematic calibration errors between the two sets of detectors \citep[see e.g., the discussion on the Solar Black window coating of STIX in][]{Stiefel_2025}, we introduce a binding parameter C which multiplies the total photon model when jointly fitting the two spectra. This parameter is set to one for the BKG detector spectrum, and is allowed to vary for the imaging detector spectrum, leaving it the freedom to adapt for systematic differences between the two\footnote{See the Appendix for some more discussion on the binding parameter C.}.

    The following energy range limits should be used for the joint fitting of STIX detectors: The lower limit of the imaging detectors is 12 keV. The upper limit depends on the statistics and, therefore, on the flare size. For the lower limit of the BKG detector, 6 keV should be used, see also the discussion in \citet{Stiefel_2025}. The top limit depends again on the flare size. Typically for large X-class flares around the peak time, there is sufficient statistics up to 25 keV. 

   A full demonstration how to fit STIX data using SUNKIT-SPEX including joint fitting is available online\footnote{\stixdemo}.

\subsection{Scientific use of joint fitting with STIX}
    In Figure \ref{Fig: Single Fit Imaging}, the spectrum for a time range with an inserted attenuator of an estimated X5-flare\footnote{The estimate is based on the method described in \citet{Stiefel_2025}, see also Section \ref{Chapter: Application, Procedure}.} on the 3rd of January 2023, measured by the imaging detectors of STIX, is shown. The top and bottom panels show the same measurement and fitted model, except that for the top panel, only one thermal model is used, and in the lower panel, two thermal components are used. In the residual panels, we present the reduced $\chi^2$ for each fit with a 3\% systematic error, which is commonly used for STIX. The reduced $\chi^2$ does not distinctly favor one of the fitting models. Looking at the residuals of the fit in the top panel, we see that it exhibits slightly worse systematics. Around 20 keV, which marks the transition between the thermal and the nonthermal model, the residuals show a systematic increase. In contrast, the residuals in the bottom panel show less variation. However, based on the fitting results alone, it is not immediately clear which fit is preferable. Additionally, it is challenging to constrain the lower temperature thermal component due to the limited number of data points.  
    In a side note, the offset in the residuals around 60 keV are due to an incomplete grid transmission implementation in the STIX GSW due to the tungsten K-edge. This will be updated in the next calibration round of STIX.

   In the next step, we include the spectrum measured by the BKG detector. In Figure \ref{Fig: Joint Fit} we show the BKG detector spectrum to the left and the imaging detector spectrum to the right for the same flare and same time interval as in Fig. \ref{Fig: Single Fit Imaging}. Using the BKG and imaging detector spectra for joint fitting, we do the same as before: In the top panel only one thermal component is included into the photon model, in the bottom panel two thermal components are used. The fitted temperature of the isothermal model is similar to the Fig. \ref{Fig: Single Fit Imaging} fit. However, the lower energy residuals and the systematics in the residuals around 20 keV indicate that this fit is inadequate. Fitting two thermal components clearly represents the observations better. This observation is supported by the reduced $\chi^2$ shown in the residual plots of the imaging detectors. The $\chi^2$ values are calculated using all energy bins within the fitting range.

    In this example, we have demonstrated the advantage of the joint fitting method for STIX flares with inserted attenuator. For a more complete understanding of the thermal emission in these flares, joint fitting is needed. For a rough estimate of the nonthermal parameters, using only the imaging detectors might be adequate. However, having a more precise constrain on the thermal components(s) will enhance and refine the nonthermal model. Therefore, we recommend using joint fitting for any flare with an inserted attenuator.

    \subsection{Advantages of joint fitting compared to the iterative method}

    \begin{figure*}
   \centering
        \includegraphics[width=18cm]{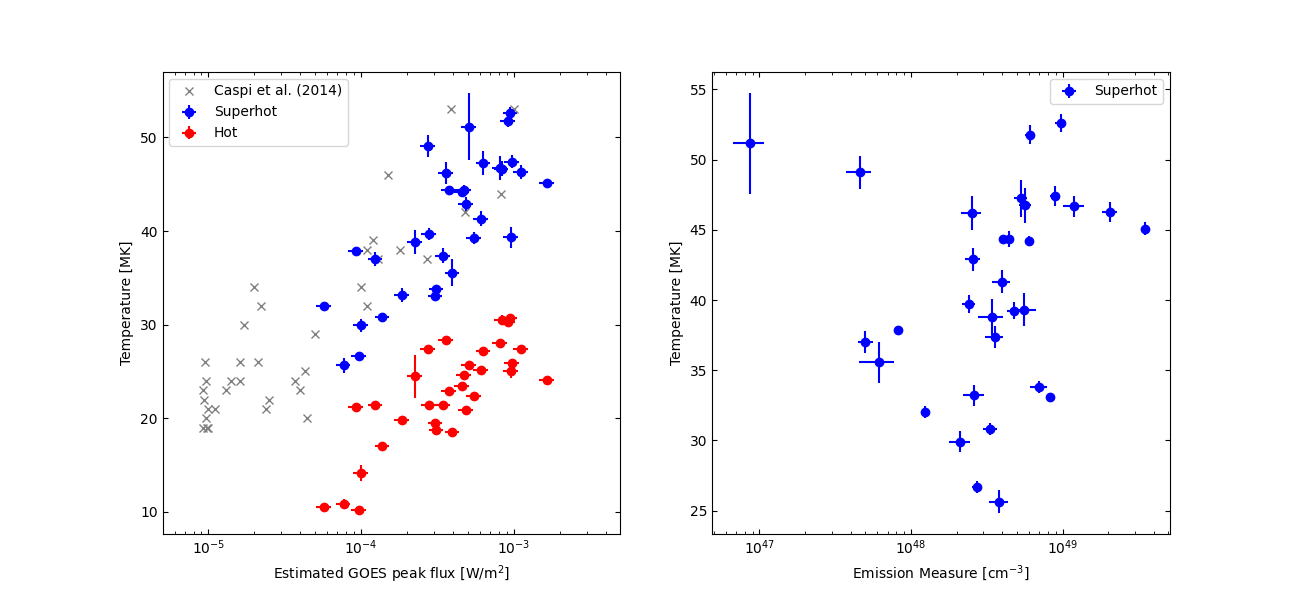}
   \caption{Results of the statistical analysis using 32 STIX flares. In the left plot, the temperature of the thermal emission as a function of the GOES flare class is shown. The x-axis is the GOES peak flux estimated on the low-energy counts from STIX, and the y-axis gives the temperature of the thermal emission in MK. The blue and red dots represent the spectral fitting results for the superhot and the hot thermal component from 32 large STIX flares observed over the past three years (2022-2024). The grey crosses are the results from the statistical analysis in \citet{Caspi_2014} with RHESSI data. The right plot shows the temperature as a function of the emission measure of the superhot component.}
    \label{Fig: Goes vs temperature}
    \end{figure*}

    We have introduced a new method for jointly fitting two spectra obtained from different sets of STIX detectors, and we have demonstrated that this approach improves thermal spectral fitting for attenuated flares. A similar approach was described in \citet{Stiefel_2025}, where the two spectra (BKG detector and imaging detectors) were fitted iteratively using OSPEX. In the iterative method, only one spectrum at a time can be fitted. This requires manually switching back and forth between the two spectra to identify a single fit that fits both spectra as closely as possible.

    A main advantage of the joint fitting is the automatic and simultaneous contribution of both spectra to constrain the parameters of the models. In contrast, in the iterative method, the model is fitted manually to both spectra. Consequently, the iterative method is more error-prone, less objective, and takes up more time. 

   Further, SUNKIT-SPEX can run an MCMC analysis. This allows us to map out the parameter posterior space to obtain the parameter confidence intervals. An example of a corner plot from an MCMC analysis is given in Appendix \ref{Appendix: MCMC} for the fit shown in Fig. \ref{Fig: Joint Fit} (b).

\section{Application: Statistical analysis of the superhot component} \label{Chapter: Application}

   \begin{table}
      \caption[]{Date and estimated GOES flare class of the 32 STIX flares.}
      \vspace{-1.8em}
         \label{Tab: Flare candidates}
         $$
         \vspace{-0.8em}
         \begin{array}{|c|c|c|}
             \hline
             & &   \\
             \mathrm{Date} & \mathrm{Integrated\;time\;range} & \mathrm{GOES\;class} \\
             \mathrm{[yyy-mm-dd]} & \mathrm{[UTC\;(Earth)]} & (\mathrm{Estimated})\\
               & &   \\\hline
              & &   \\
             2022-03-30 & 17:34:30 - 17:34:40 &  \mathrm{X1.4\pm0.2} \\
             2022-04-13 & 13:00:10 - 13:00:20 &  \mathrm{X1.9\pm0.2} \\
             2022-04-20 & 03:54:50 - 03:55:00 &  \mathrm{X2.7\pm0.3} \\
             2022-09-29 & 11:55:45 - 11:55:55 &  \mathrm{X4.6\pm0.5} \\
             2023-01-03 & 06:26:50 - 06:27:00 &  \mathrm{X4.8\pm0.5} \\
             2023-03-29 & 02:30:50 - 02:31:00 &  \mathrm{X1.2\pm0.1} \\
             2023-07-16 & 04:30:30 - 04:30:40 &  \mathrm{X5.5\pm0.6} \\
             2023-07-16 & 13:30:40 - 13:30:50 &  \mathrm{X4.7\pm0.5} \\
             2023-07-17 & 00:40:50 - 00:41:30 &  \mathrm{X9.2\pm1.0} \\
             2023-09-08 & 11:17:10 - 11:17:20 &  \mathrm{X3.4\pm0.4} \\
             2023-09-11 & 22:25:20 - 22:25:30 &  \mathrm{X3.1\pm0.3} \\
             2023-12-31 & 21:41:10 - 21:41:20 &  \mathrm{X4.0\pm0.4} \\
             2024-02-02 & 13:26:20 - 13:26:30 &  \mathrm{X2.8\pm0.3} \\
             2024-02-22 & 22:29:10 - 22:29:20 &  \mathrm{X6.3\pm0.7} \\
             2024-03-28 & 15:55:20 - 15:55:30 &  \mathrm{M7.7\pm0.8} \\
             2024-03-28 & 20:39:00 - 20:39:10 &  \mathrm{M9.7\pm1.0} \\
             2024-03-30 & 21:13:40 - 21:13:50 &  \mathrm{X1.0\pm0.1} \\
             2024-04-03 & 03:14:50 - 03:15:00 &  \mathrm{M9.2\pm1.0} \\
             2024-04-03 & 16:44:00 - 16:44:10 &  \mathrm{M5.7\pm0.6} \\
             2024-05-14 & 16:48:10 - 16:48:20 &  \mathrm{X11.1\pm1.2} \\
             2024-05-14 & 02:06:10 - 02:06:20 &  \mathrm{X2.2\pm0.2} \\
             2024-05-15 & 20:37:20 - 20:37:30 &  \mathrm{X9.5\pm1.0} \\
             2024-05-15 & 08:16:40 - 08:16:50 &  \mathrm{X9.7\pm1.1} \\
             2024-05-16 & 13:25:20 - 13:25:30 &  \mathrm{X3.8\pm0.4} \\
             2024-05-17 & 12:27:50 - 12:28:00 &  \mathrm{X8.1\pm0.9} \\
             2024-05-20 & 05:16:00 - 05:16:10 &  \mathrm{X16.5\pm1.8} \\
             2024-06-17 & 10:20:30 - 10:20:40 &  \mathrm{X3.1\pm0.3} \\
             2024-07-22 & 23:51:50 - 23:52:00 &  \mathrm{X9.6\pm1.1} \\
             2024-07-27 & 04:33:00 - 04:33:10 &  \mathrm{X8.3\pm0.9} \\
             2024-08-01 & 20:52:20 - 20:52:30 &  \mathrm{X3.6\pm0.4} \\
             2024-10-01 & 22:11:50 - 22:12:00 &  \mathrm{X5.1\pm0.6} \\
             2024-10-03 & 12:16:00 - 12:16:10 &  \mathrm{X6.1\pm0.7} \\
             & &   \\
            \hline
         \end{array}
         $$ 
         \tablefoot{For each flare of the 32 used in Section \ref{Chapter: Application} the date, integration time for the spectral fit (in Earth time) and the estimated GOES flare class is given.}
   \end{table}

    \citet{Caspi_2014} analyzed 37 RHESSI flares with GOES classes between C9 and X10. The statistical analysis showed that flares with a GOES class > X1 do have a superhot component with temperatures above 30 MK. Further, \citet{Caspi_2014} reported a correlation between the GOES class and the temperature of the superhot component. We can confirm these results using 32 STIX flares and the joint fitting method described in Section \ref{Chapter: Method}.

\subsection{Method} \label{Chapter: Application, Procedure}

    For the analysis, we choose 32 STIX flares accordingly: Using a list of STIX flares with attenuator movement (to date (June 2025) around 150 flares), we sorted them by counts in the 15-25 keV quicklook channel. From this dataset, we selected the 32 largest flares. These occurred between March 2022 and October 2024, with estimated GOES classes ranging from M5 to X16. The estimation of the GOES classes is based on the method described in \citet{Stiefel_2025}, together with the suggested 11\% error on the estimate. Table \ref{Tab: Flare candidates} summarizes the 32 flares used in our analysis.

    We carried out joint spectral fitting for each of the 32 flares at the temperature peak time, integrated over a 10s interval. The temperature peak time was determined by fitting flare spectra at several time steps around the thermal peak using an isothermal model. The time step with the highest temperature was selected as the temperature peak time. As described in Section \ref{Chapter: Method}, we jointly fitted two thermal components, one nonthermal component, and an albedo component to the spectra obtained from the BKG detector and the imaging detectors.

\begin{figure}
    \centering
    \includegraphics[trim=00 00 00 00, clip, width=0.45\textwidth]{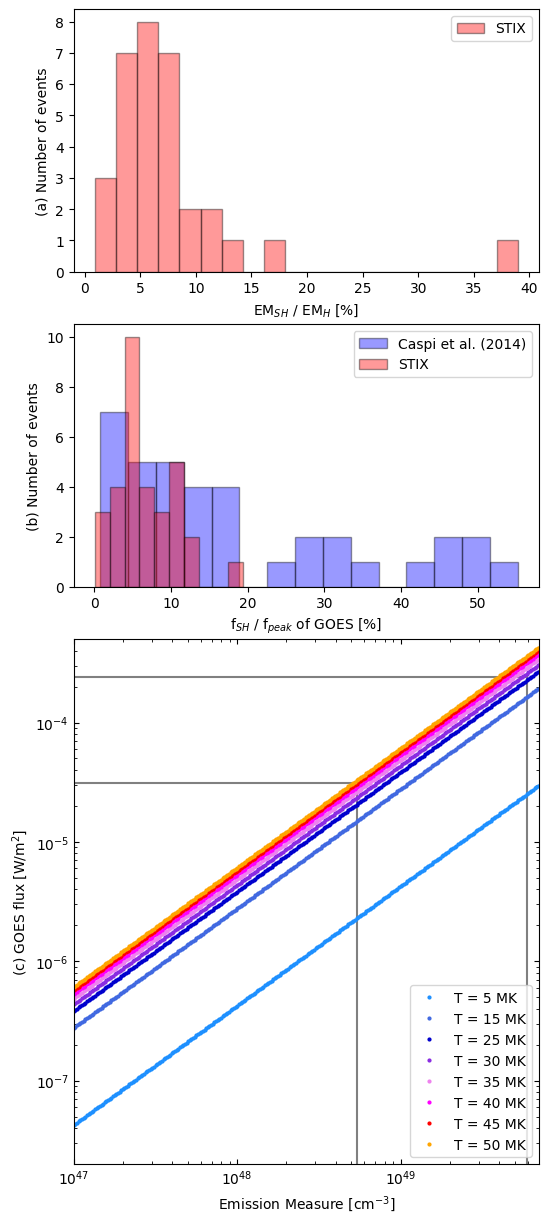}
    \caption{Panel (a) shows the histogram plot of the ratio in percentage between the superhot EM and the hot EM using the 32 STIX flares. Panel (b) gives the ratio in percentage between the estimated GOES flux of the superhot component and the GOES peak flux (= GOES flare class). Panel (c) shows the GOES flux as a function of the EM. The different colors correspond to different temperatures, from blue to orange, as the temperature increases.}
    \label{Fig: EM Ratio and GOES}
\end{figure}

\subsection{Results and discussion}

    In Figure \ref{Fig: Goes vs temperature} to the left, we show the results from the spectral analysis for the temperatures of the hot and the superhot thermal components as a function of the estimated GOES class. In grey, the RHESSI flares analyzed by \citet{Caspi_2014} are included for comparison. Unlike this paper, \citet{Caspi_2014} only used an isothermal model to fit the thermal component. This is justified using the assumption that the thermal emission measured by RHESSI is largely dominated by the superhot component. Regardless of this difference, the new results using STIX data align well with the findings from RHESSI. Additionally, STIX also provides results on the parameters of the hot thermal model. The temperatures of the superhot component are generally in the range $\geq$ 30 MK. Moreover, STIX data also reveal a trend that flares with a higher GOES class reach higher temperatures in the superhot component. To quantify this relationship, we calculated the Pearson correlation coefficient, which is a measure of linearity between two datasets. For this calculation, we used the logarithm of the estimated GOES flux and the temperature of the superhot component. We obtained a correlation coefficient of 0.77. with notable significance based on a dataset of 32 flares. Loosely speaking, higher GOES class flares therefore reach higher temperatures in the superhot component. However, we note that there is also a significant spread of temperatures per GOES class, highlighting the complex nature of solar flares. In their study, \citet{Caspi_2014} reported a Pearson correlation coefficient of 0.88, indicating a stronger correlation. One notable difference between the dataset developed in this paper and that of \citet{Caspi_2014} is the range of GOES classes. When we recalculate the Pearson correlation coefficient, including only flares >M5 from the \citet{Caspi_2014} dataset, we obtain a value of 0.79, which is closer to the value derived in this study.

    In several earlier studies, the relationship between the GOES class and plasma temperature was examined. These studies can be categorized into three distinct groups: 1. Studies that used the GOES temperatures \citep[e.g., ][]{Ryan_2012, Caspi_2014, Warmuth_2016}. 2. Studies that used the HXR instrument temperatures derived from a single thermal fit \citep[e.g., ][]{Battaglia_2005, Warmuth_2016}. 3. Studies that focused on the superhot temperatures \citep[e.g., ][]{Caspi_2014}. The model commonly used between the GOES class and the plasma temperature is given by
    \begin{equation}
        T\;=\;a\log_{10}(F_G) + b 
    \end{equation}
     where T is the temperature, and $F_G$ denotes the GOES flux. We fitted the curve to our data, and determined the values $a=11.13\pm0.03$ and $b=60.5\pm0.3$ for the hot, and $a=13.1\pm0.1$ and $b=84\pm1$ for the superhot component, respectively. The result for the superhot component is very similar to the findings of \citet{Caspi_2014}, which reported a slope of $14$. The slope we determined for the hot component is slightly smaller. \citet{Caspi_2014} used the GOES temperature to determine the slope of the low temperature component, yielding a result of $4.6$, which is significantly lower than what we report here. Our findings align more closely with those of \citet{Warmuth_2016} with a slope of $9.5$, reinforcing their conclusion that the results depend on whether the temperature is derived from GOES or from a HXR instrument like STIX or RHESSI. This discrepancy is likely due to the different sensitivities of the instruments to plasma temperatures \citep[similar to what was discussed in][]{Stiefel_2025}.
     
     It is essential to note that each study is biased by the selection of flares. The current study focuses exclusively on the largest flares, while, for example, \citet{Warmuth_2016} analyzed flares ranging from A to X classes. For a more complete understanding, future research should involve a larger sample of flares spanning various GOES classes. The method presented in this paper can aid in the inclusion and analysis of the largest flares.
    
    For the smaller STIX flares in this study (M5-X1), the STIX spectra did not always give a conclusive answer if a double thermal or an isothermal fit is better. This is motivated by examining the fits, the residuals, and the reduced $\chi^2$. For instance, SOL2024-03-28T1500, which is an estimated M7.7 class flare, has reduced $\chi^2$ values of 2.16 and 2.25 for single or double thermal fit, respectively. Unlike the example presented in Fig. \ref{Fig: Joint Fit}, this case does not provide a conclusive answer regarding which photon model better represents the data. This implies that either the superhot (or the hot) component is not detectable anymore within the dynamic range of STIX, or that the superhot component is absent in these smaller flares. Similar can be seen in the \citet{Caspi_2014} dataset. Most of the flares smaller than X1 have temperatures below the 30 MK superhot range.

    In Figure \ref{Fig: Goes vs temperature} to the right, the temperature of the superhot component is plotted as a function of its emission measure (EM). There is no linear correlation visible between the two values, which is also confirmed by the Pearson correlation coefficient (in log-linear scale) with a value of 0.11. As discussed in \citet{Caspi_2014}, if the high temperature component in the spectral fitting is an artefact of the data, a linear anti-correlation between the EM and the temperature would be expected. The lack of correlation confirms the fact that the high temperature component is a real physical component of the spectrum.

    \subsubsection{Contribution of the superhot component in soft X-rays}

    HXR instruments such as STIX or RHESSI are designed to observe the hottest plasma in solar flares, which makes them good candidates for observations of the superhot component. Nevertheless, we would like to qualitatively discuss whether other instruments, such as GOES or the X-ray Telescope \citep[XRT;][]{Golub_2007} on Hinode, can detect signals of the superhot component. The temperature response of both GOES and XRT flattens at higher temperatures \citep{Golub_2007}. This means that in the high temperature range, the measured signal is more strongly influenced by EM than by temperature (see also Fig. \ref{Fig: EM Ratio and GOES} (c)). Therefore, we aim to compare the EMs of the hot and the superhot components from the STIX analysis.

    The ratios of the superhot EM to the hot EM for the 32 STIX flares are given in the histogram plot in Fig. \ref{Fig: EM Ratio and GOES} (a). For most of the flares, the superhot EM is around 5-10\% of the hot EM, even though there is a spread showing the individual thermal nature of each flare. We point out that these results are for times at the temperature peak of the flare. In general, the contribution of the superhot component decreases compared to the hot component during the decay phase of a flare \citep[e.g., ][]{Caspi_2010}.

    With the dynamical range of XRT, plasma with EM ratios of 1:10 should be observable. Therefore, we expect to have a chance of detecting the superhot component for certain flares and during specific time points with XRT. However, this is strongly dependent on the shape and overall temperature distribution of the flares. 
    
    To distinguish emissions from the superhot and hot components, HXR instruments, such as STIX, would greatly support the analysis. A new approach to image the individual thermal components using STIX will be the focus of our next paper. Multi-instrument observations of superhot components would then contribute to our understanding of the origins of these superhot features.

    To provide a more detailed analysis of the GOES measurements, we aim to estimate the contribution of the superhot component in the total GOES signal. To achieve this, we estimate the GOES flux f$_{SH}$ of the superhot component based on the determined temperature and EM\footnote{This calculation is performed using the routine "goes\_fluxes.pro" in IDL using the satellite keyword 12 and 16 for RHESSI and STIX flares, respectively.}. We then compare this value with the total GOES peak flux $f_{peak}$. The results are presented in Fig. \ref{Fig: EM Ratio and GOES}, panel (b), which includes the flares studied in this paper as well as those from the \citet{Caspi_2014} paper. For the majority of the flares analyzed, the superhot component accounts for 1-10 \% of the GOES peak flux. 
    
    We would like to note that we are comparing two different time points in the histogram: the GOES peak and the temperature peak time. This simplification is necessary because approximately 50\% of the STIX flares do not have any actual GOES measurements, which means we can only estimate the GOES peak flux. Additionally, this approach allows us to include the \citet{Caspi_2014} flares. Since the GOES flux f at the temperature peak time is always $f \leq f_{peak}$, the percentages shown in the histogram plot represent lower limits. Nonetheless, we can argue that the simplification is reasonable when comparing with panel (a) of Fig. \ref{Fig: EM Ratio and GOES}. In the higher temperature range, the EM is the critical factor for the GOES signal (see panel (c)). Taking one example of a 1:10 ratio of the EM for the superhot to the hot component, we can estimate the GOES signal for each component individually (represented by the grey lines in panel (c)). The GOES signal of the superhot is roughly 11\% of the GOES signal of the hot component, confirming that the superhot component can contribute a few percent to the GOES signal.

    The spread of the \citet{Caspi_2014} flares in Fig. \ref{Fig: EM Ratio and GOES}, panel (b) is larger than the STIX dataset. We examined the flares with percentages above 30\%. All of these flares have a GOES class below M5. It is possible that for these flares, the superhot component is not distinctly visible due to limitations of instrumental dynamical ranges, or does not exist. Therefore the signal measured by RHESSI would primarily reflect the hot component with higher EM values.

    \subsubsection{Comparing the thermal energy content} 
    
    An essential parameter to compare between the hot and the superhot components is the thermal energy, which can be calculated using the following equation
    \begin{equation}
        E_{th}\;=\;3k_BT\sqrt{EM\cdot V},
    \end{equation}
    where V represents the volume, EM is the emission measure, and T is the temperature. To estimate the volume, we need to be able to image the two components separately, which will be the topic of our future paper. Since we can not directly calculate the thermal emission, we will analyze the ratio given by: 
     \begin{equation}
        \frac{E_{SH}}{E_{H}}\;=\;\frac{T_{SH}}{T_{H}}\sqrt{\frac{EM_{SH}}{EM_{H}}}\sqrt{\frac{V_{SH}}{V_{H}}}\;\approx\;\frac{2}{\sqrt{10}}\sqrt{\frac{V_{SH}}{V_{H}}}.
    \end{equation}
    For the temperature and the EM ratios, we took average values from the analysis of this paper. Based on \citet{Caspi_2015}, we can assume that the volume of the superhot component is smaller than or equal to the volume of the hot component. If the volume is equal, the thermal energy of the superhot can be up to $\thicksim$ 2/3 of the thermal energy of the hot component. Even for a 1:10 ratio between the volumes, the thermal energy of the superhot is still 20\% of the thermal energy of the hot component, indicating that both thermal components contain a significant amount of the energy in solar flares. Therefore, understanding the individual thermal components better by using joint fitting for STIX and two thermal components can also improve the understanding of the overall flare energy budget.
    
\section{Discussion and Conclusion} \label{Chapter: Conclusions}

    In this paper, we presented a new method for jointly fitting the spectrum of the BKG detector and the imaging detectors of STIX using SUNKIT-SPEX. This approach enhances thermal fitting, particularly when distinguishing between different plasma populations, such as the hot and superhot thermal components. The BKG detector constrains the hot component, while the imaging detectors constrain the nonthermal, and both detector combinations constrain the superhot component. A key advantage of this joint fitting method, compared to other studies that also used joint fitting for multiple spectra, is that both spectra are observed from the same detector type. This means the knowledge of the calibration is the same, resulting in only minor differences between the two spectra (as detailed in Appendix \ref{Appendix: C-Value}).

    We demonstrated the effectiveness of the new method in a statistical analysis of large flares (> M5) observed by STIX over the last three years. Our results agree with previous studies on the thermal emission of X-class flares \citep[e.g., ][]{Caspi_2014}. Our results confirm that large flares are better represented by two thermal components that correspond to two distinct plasma populations, rather than by a single thermal model. 

\subsection{Multithermal nature of solar flares}

    In this paper, we focused on the discussion of isothermal or double thermal models. In both cases, each plasma population is described by a single temperature. However, associating plasma with a single temperature does not fully capture the complexity of solar flares \citep[e.g., ][]{Warren_2013}. Instead, this temperature is an average value of all the temperatures within the plasma, weighted by the temperature response of the instrument used for the observation. In the extreme ultraviolet (EUV), differential emission measure (DEM) analyses are commonly used \citep[e.g., ][]{Hannah_2012}. However, in the SXR and HXR range, and therefore in the high-temperature ranges, DEM analysis is less common and not as well-constrained. 

    Nevertheless, DEM analyses in the SXR/HXR range suggest that the simplified approach of fitting two single temperatures is a reasonable approximation. The DEM analysis conducted by \citet{McTiernan_1999} revealed two individual temperature distributions for certain flares - one in the lower temperature range and another in the higher. Additionally, a more recent study by \citet{Mithun_2022} demonstrated that the spectra measured by the Solar X-ray Monitor \citep[XSM; ][]{Vadawale_2014} are best fitted with a two-peaked differential emission distribution, indicating the presence of two distinct thermal populations. Our results support these DEM reports.

    As a next step for a more detailed analysis of the thermal emission in solar flares, STIX analysis will benefit from joint analysis with a SXR instrument, such as XSM or the High Energy L1 Orbiting X-ray Spectrometer \citep[HEL1OS; ][]{Sankarasubramanian_2017_Hel1os} on the Aditya-L1 mission \citep{Tripathi_2023_Aditya} to understand the lower temperature ranges better, similar to what has been done by \citet{Nagasawa_2022} for RHESSI.

\subsection{Future HXR space missions}

    In the context of future HXR instruments, we discuss the advantages of a "BKG detector" setup, a single detector with well-defined openings and without a movable attenuator. Future HXR instruments will also need attenuators to prevent pile-up. The BKG detector setup, as STIX uses it, fills in the gap and provides complementary information on the missing energy ranges during times when the attenuator is in use. We have demonstrated that these energy ranges are crucial diagnostic tools for understanding the different thermal plasma populations in large solar flares, an area with several unresolved questions. One advantage of the BKG detector setup compared to the RHESSI attenuator system is the more straightforward calibration of the setup. Calibrating an attenuator with non-uniform thickness, such as RHESSI used it, is more complex and prone to errors. 

    Furthermore, we propose the idea of a "BKG detector setup" as a flare monitor for future space weather missions. As discussed in \citet{Stiefel_2025}, this setup enables continuous X-ray flux monitoring across a broad range of GOES classes. The STIX BKG detector design is driven by the varying radial distance of Solar Orbiter from the Sun. For Earth-based missions, which have a fixed distance to the Sun, the design can be adapted. By using fixed entrance windows ("fixed attenuators") with multiple different-sized openings, we can monitor both thermal and nonthermal emission. Additionally, controlling the individual pixels by turning them on and off will help to reduce saturation.

\subsection{Conclusion}
     In conclusion, this paper presents a new method of joint fitting between two different detector configurations of STIX. For large, attenuated STIX flares, this approach improves the spectral fitting of the thermal emissions.

\begin{acknowledgements}
      \em{We} thank the referee for the constructive feedback which helped to improve the paper significantly. Solar Orbiter is a space mission of international collaboration between ESA and NASA, operated by ESA. The STIX instrument is an international collaboration between Switzerland, Poland, France, Czech Republic, Germany, Austria, Ireland, and Italy. NB acknowledges support from the UK’s Science and Technology Facilities Council (STFC) doctoral training grant (SST/X508391/1). MS thanks the Institute for Data Science at FHNW for their continued support during her doctoral studies. We want to thank Dr. Marina Battaglia for her feedback and valuable input on the paper. 
\end{acknowledgements}

\bibliographystyle{aa}
\bibliography{Joint_Fitting}

\appendix
\section{MCMC example plot} \label{Appendix: MCMC}
\begin{figure*}
    \centering
        \includegraphics[width=18cm]{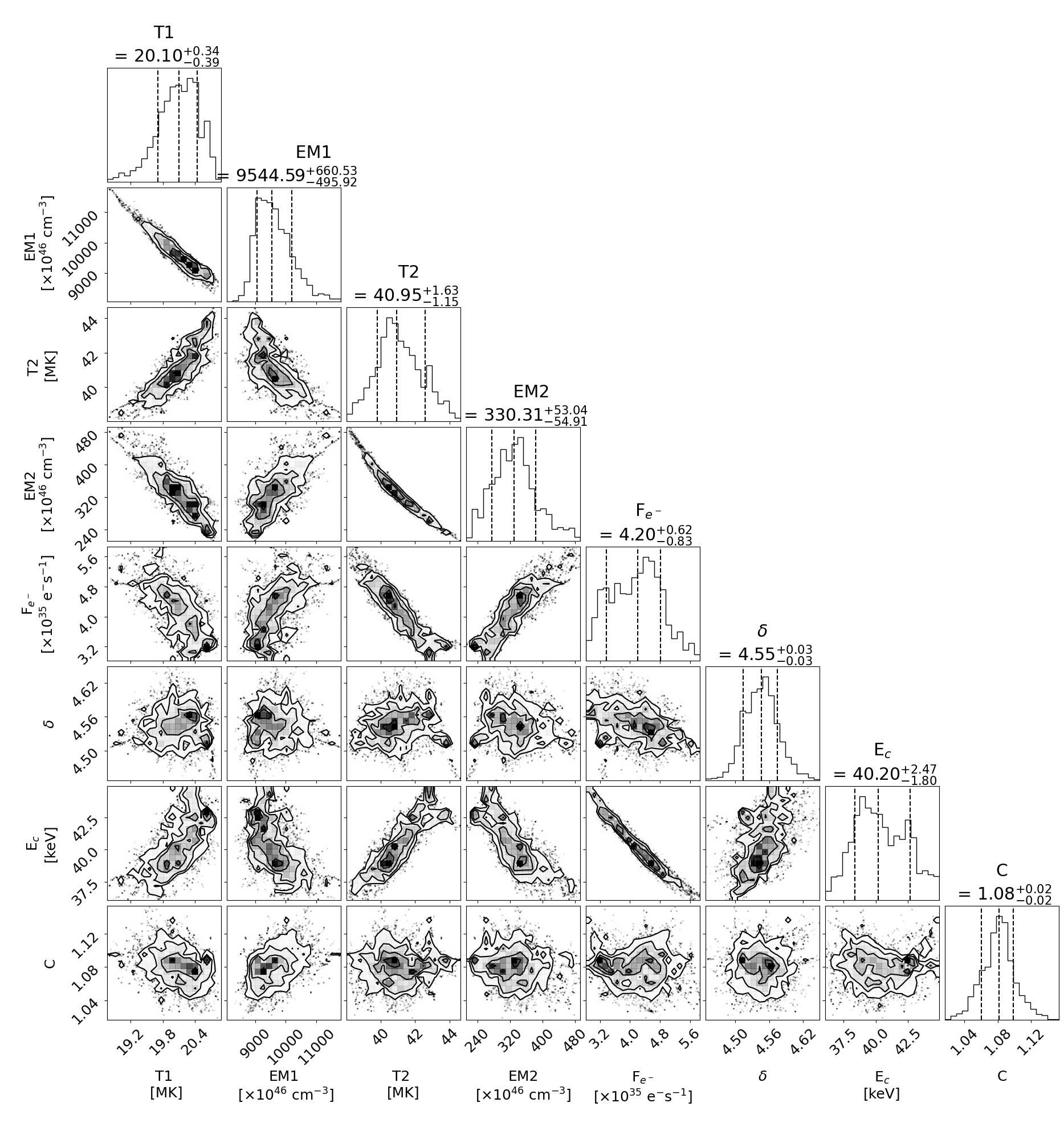}
    \caption{MCMC analysis of the fit shown in Fig. \ref{Fig: Joint Fit} (b). The diagonal plots show the histogram plots of the individual parameters fitted in the photon model. The other plots show the correlations between two different parameters.}
    \label{Fig: MCMC Example}
    \end{figure*}

    In SUNKIT-SPEX, a MCMC analysis can be run on the spectral fit. In Figure \ref{Fig: MCMC Example} the results of the MCMC analysis of the fit in Fig. \ref{Fig: Joint Fit} (b) are shown. All parameters are fitted to both spectra except for the binding parameter C which is only fitted to the imaging detectors and compensates for systematic differences between the BKG detector and the imaging detectors. In Appendix \ref{Appendix: C-Value} a short discussion on the binding parameter C is included. 

\section{Binding parameter C} \label{Appendix: C-Value}
\begin{figure}
        \centering
        \includegraphics[width=9cm]{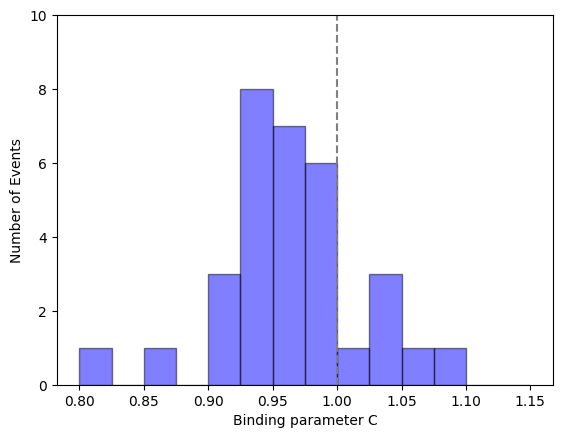}
        \caption{Histogram plot of the binding parameter C for the 32 flares used in this study.}
        \label{fig: C-value Analysis}
    \end{figure}

     In Figure \ref{fig: C-value Analysis} a histogram plot is shown for the binding parameter C of the 32 attenuated flares analyzed in this study. The binding parameter is fitted to the imaging detectors and set to one for the BKG detector. Therefore a C value below 1 means that the imaging detectors had systematically less flux compared to the BKG detector and a C value above 1 the opposite. Most of the flares show a value below 1 with a peak around 0.95. This means a smooth-out 5\% systematic difference over all energy ranges between the two detector configurations. As a wide energy range is used in fitting the binding parameter, the factor represents an averaged systematic error in the currently available calibration, but we cannot attribute it to a specific effect.

\end{document}